\documentclass[twocolumn,prl,showpacs,aps,floatfix]{revtex4}

\usepackage{graphicx}
\usepackage{amssymb}
\usepackage{bm}
\usepackage[usenames]{color}
\newcommand{\be}{\begin{equation}}
\newcommand{\ee}{\end{equation}}
\newcommand{\br}{\begin{eqnarray}}
\newcommand{\er}{\end{eqnarray}}

\begin{document}

\title{Minimal set of local measurements and classical communication 
for two-mode Gaussian state entanglement quantification}
\author{Luis F. Haruna, Marcos C. de Oliveira, and Gustavo Rigolin}
\affiliation{Instituto de F\'\i sica ``Gleb Wataghin'', Universidade 
Estadual de Campinas, 13083-970, Campinas, S\~ao Paulo, Brazil.}

\begin{abstract}
We develop the minimal requirements for the
complete entanglement quantification of an arbitrary two-mode bipartite
Gaussian state via local measurements and a classical
communication channel. The minimal set of measurements is
presented as a reconstruction protocol of local covariance
matrices and no previous knowledge of the state is required but
its Gaussian character. The protocol becomes
very simple mostly when dealing with Gaussian states transformed
to its standard form, since  photocounting/intensity measurements define the 
whole set of entangled states. In addition, conditioned on some
prior information, the protocol is also useful for a complete
global state reconstruction. 
\end{abstract}

\maketitle

Quantum communication protocols extend the information theoretical
notion of channel \cite{Thomas} to the quantum domain by
incorporating non-local entangled states. Those channels are
generated by the preparation of a pair (or more) of quantum systems in an
entangled state, which are then separated to establish non-local
correlations \cite{EPR}, allowing several communication tasks
otherwise unattainable via classical channels \cite{nielchu}.
However, for most of the quantum protocols to work properly
(deterministically) one has first to be able to prepare maximally
pure entangled states and then to guarantee that those states stay
pure or nearly pure during all the processing time.
An important problem then 
arises in this whole process: One has to check the ``quality''
(the amount of entanglement and purity) of the quantum channel,
while usually the only available tools for that are local
measurements (operations) and one (or several) classical channel.

The quest for an optimal and general solution for this problem has
generated a vast literature on the
characterization of entangled states under local operations and
classical communication (LOCC), 
either for
qubits \cite{nielsen} or 
for continuous variable systems of the Gaussian type \cite{eisert1,eisert2}.
Gaussian states (completely
described by 
up to second order moments) are particularly important since they
can be easily generated with radiation field modes.
Moreover, operations that keep the
Gaussian character (so-called Gaussian operations) are given by
the transformations induced by linear (active and passive) optical
devices (beam-splitters, phase-shifters, and squeezers)
\cite{eisert2}. A particular 
result for this kind of state is that it is impossible to distill
entanglement out of a set of Gaussian states through Gaussian
operations \cite{eisert3}.

Assuming one is left with only Gaussian local operations and a
classical channel (GLOCC), how is it possible to infer the quality
of a quantum channel in use? For a two-mode Gaussian state
one possibility is to 
access directly the entanglement properties of the
system after a proper manipulation of the two modes
\cite{marcos2,rigolin}. This procedure requires, however, that the
two parties (modes) be recombined in a beam-splitter (non-local
unitary operation) in which their entanglement content
are transferred to 
local properties of one of the output modes. Another possible way
is to completely reconstruct the bipartite quantum system, a
resource demanding task \cite{laurat} which
also requires global operations here
forbidden.

In this Letter we demonstrate 
a minimal set of GLOCC to completely 
quantify the entanglement of a two-mode Gaussian state. As a
bonus of this procedure one can also assess the purity of the
Gaussian state and, for some particular classes of states,
reconstruct the bipartite covariance matrix.
The protocol consists mainly in the attainment, via local
measurements, of all the symplectic invariants that allows, for example,
one to test the separability of the system, to know its
P-representability properties, and to quantify its entanglement
content.
We also show that for a particular class of Gaussian states
belonging to the set of symmetric Gaussian states \cite{marcos1},
the Einstein-Podolsky-Rosen (EPR) states and general mixed
squeezed states , the protocol becomes
straightforward due to the relative easiness one obtains the
correlation matrix elements from local measurement outcomes.
Moreover, since P-representability and separability for these kind of states
are equivalent, we show that for 
two-mode thermal squeezed states with internal noise \cite{daffer}
it is possible to decide whether or not they are separable 
via local photon number measurements.

A two-mode Gaussian state $\rho_{12}$ is characterized by its
Gaussian
characteristic function
$C({\bm\alpha})=e^{-\frac12{\bm{\alpha}^\dagger}{\bf
V}{\bm{\alpha}} }$, where $\bm{\alpha}^\dagger=\left(\alpha_1^*,
\alpha_1, \alpha_2^*, \alpha_2\right)$ are complex numbers and
$a_1$ ($a_1^\dagger$) and $a_2$ ($a_2^\dagger$) the annihilation
(creation) operators for parties 1 and 2, respectively
\cite{comment1}. The covariance matrix \textbf{V} describing all
the second order moments $V_{ij}=(-1)^{i+j}\langle v_i v_j^\dagger
+ v_j^\dagger v_i \rangle/2$, where
$(v_1,v_2,v_3,v_4)=(a_1,a_1^\dagger,a_2,a_2^\dagger)$, is given by
\begin{displaymath} \textbf{V}=\left(
\begin{array}{cc}
\textbf{V}_1 & \textbf{C} \\
\textbf{C}^\dagger & \textbf{V}_2 \end{array} \right)
=\left(\begin{array}{cccc}
n_1+ \frac{1}{2} & m_1 & m_s & m_c \\
m_1^* & n_1+\frac{1}{2} & m_c^* & m_s^* \\
m_s^* & m_c & n_2+\frac{1}{2} & m_2 \\
m_c^* & m_s & m_2^* & n_2+ \frac{1}{2} \\ \end{array} \right).
\end{displaymath}
 $\textbf{V}_1$ and $\textbf{V}_2$ are local Hermitian matrices 
while \textbf{C} is the correlation between the two parties.
 Any covariance matrix must be positive semidefinite
$\textbf{V}\geq\mathbf{0}$ and the generalized
uncertainty principle, $\textbf{V}+(1/2)\textbf{E}\geq\mathbf{0}$, where 
${\bf E}=\text{diag}(\mathbf{Z},\mathbf{Z})$ and  
${\bm{Z}}=\text{diag}(1,-1)$, must hold \cite{englert}.

From local measurements on both modes of $\rho_{12}$, either
through homodyne detection (see \cite{grangier} and references
therein) or alternatively by employing single-photon detectors
\cite{fiurasek2}, the local covariance matrices $\textbf{V}_1$ and
$\textbf{V}_2$ can be reconstructed. Remark that for the
reconstruction of the global matrix \textbf{V}, and therefore the
joint bipartite state, one has to obtain \textbf{C}. Obviously,
global joint measurements achieved through recombination of the
two parties in a beam-splitter followed by local homodyne
detections are forbidden. Thus one has to deal only with local
measurements whose results can be sent through classical
communication channels to the other party. As we now show, there
are minimal operations/measurements that can be performed locally on
the system to attain $|\det\textbf{C}|$ and $\det$\textbf{V}.
These quantities, together with $\det\textbf{V}_1$ and
$\det\textbf{V}_2$, will be shown to be all that one needs to
determine whether or not a two-mode Gaussian state is entangled as
well as how much it is entangled. As it will become clear, the required set 
of operations is minimal in the sense that only two local measurement 
procedures are needed - one to characterize local covariance matrices and 
another to locally assess the parity of one of the modes.

First of all let us introduce an important result \cite{haruna1}. 
Given a two-mode
Gaussian state with density operator
$\rho_{12}$ and covariance matrix $\textbf{V}$ we can define the 
Gaussian operator
$\sigma_1 = Tr_2\left\{e^{i\pi a_2^\dagger a_2}\rho_{12}\right\},$
whose covariance matrix ${\bf\Gamma}_1$ is the
Schur complement \cite{horn} of $\textbf{V}$ relative to
$\textbf{V}_2$:
\be {\bf
\Gamma}_1=\textbf{V}_1-\textbf{C}\textbf{V}_2^{-1}\textbf{C}^\dagger.
\label{schur_rel1}
\ee
The meaning of $\sigma_1$ is best appreciated through a partial 
trace in the Fock basis: 
$\sigma_1=\sum_{n_{even}} \!_2\langle n|\rho_{12}|n\rangle_2 -
\sum_{n_{odd}}\!_2\langle n|\rho_{12}|n\rangle_2=\rho_{1_{e}}-\rho_{1_{o}}$, 
being equal to the difference between Alice's mode states
conditioned, respectively, to even and odd parity measurement 
results by Bob \cite{haruna1}. 
While $\rho_{1_e}$ and $\rho_{1_o}$ are not generally Gaussian, 
$\sigma_1$ is a Gaussian operator, and ${\bf\Gamma}_1$ can be built 
with only second order moments of these conditioned states.

Now suppose that Alice and Bob share many copies of a two-mode
Gaussian state. The protocol works as follows: ({\it i}) Firstly,
in a subensemble of the copies, each party performs a set of local
measurements in such a manner to obtain the 
covariance matrices $\textbf{V}_1$ and $\textbf{V}_2$,
corresponding to the reduced operators
$\rho_1=Tr_2\{\rho_{12}\}$ and $\rho_2=Tr_1\{\rho_{12}\}$; ({\it
ii})  Then Bob informs Alice, via a classical communication
channel, 
the matrix elements of $\textbf{V}_2$; ({\it iii}) After that, for
the remaining copies, Bob performs parity measurements on his
mode, letting Alice know to which copies does that operation
correspond and the respective outcomes, i.e. even parity
(eigenvalue 1) or odd parity (eigenvalue -1); ({\it iv}) Alice
then separates her copies in two  groups, the even ($e$) and the
odd ($o$) ones. The first group ($e$) contains all the copies
conditioned on an even parity measurement on Bob's copies. The
other one ($o$) contains all the remaining copies, namely those
conditioned on an odd parity measurement at Bob's; ({\it v}) For
each group, Alice measures the respective correlation matrices
$\textbf{V}_{1e}$ and $\textbf{V}_{1o}$; ({\it vi}) Finally, she
obtains $\bf\Gamma_1$ (Eq. (\ref{schur_rel1})) subtracting the odd
correlation matrix from the even one \cite{haruna1}:
${\bf\Gamma}_1=\textbf{V}_{1e} - \textbf{V}_{1o}$. 
Remarkably, with $\textbf{V}_1$, $\textbf{V}_2$ and
${\bf\Gamma}_1$ in hand Alice
is able to
completely characterize the Gaussian state's entanglement content
as well as its purity without any global or non-local measurements.

Remembering that a two-mode Gaussian state's purity $\mathcal{P}$
is equal to $1/(4\sqrt{\det\mathbf{V}})$ \cite{adesso} and using
the identity \cite{horn}
\begin{equation}
\det\textbf{V}=\det\textbf{V}_2\det{\bf\Gamma}_1, \label{detV}
\end{equation}
Alice readily obtains the purity of the channel: $\mathcal{P}=
1/(4\sqrt{\det\mathbf{V}_2\det{\bf\Gamma}_1}).$
Her next task is to decide whether or not she deals with an
entangled two-mode Gaussian state. Using the Simon separability
\cite{simon} test she knows that it is not
entangled if, and only if,
\begin{equation}
I_1I_2 + \left( 1/4 - |I_3|\right)^2 - I_4 \geq (I_1
+ I_2)/4, \label{separabilidade}
\end{equation}
where $I_1=\det\mathbf{V_1}$, $I_2=\det\mathbf{V_2}$,
$I_3=\det\mathbf{C}$, and
$I_4=\text{tr}(\mathbf{V_1}\mathbf{Z}\mathbf{C}\mathbf{Z}
\mathbf{V}_2\mathbf{Z}\mathbf{C^\dagger}\mathbf{Z})$. 
These four quantities are the
local symplectic invariants, belonging to the $Sp(2,R) \otimes
Sp(2,R)$ group \cite{simon}, that characterizes all the
entanglement properties of a two-mode Gaussian state. Alice
already has $I_1$ and $I_2$. We must show, however, how she can
obtain $|I_3|$ and $I_4$. Since one can prove that \cite{rigolin}
\begin{equation}
I_4 = 2|I_3|\sqrt{I_1I_2}, \label{I4}
\end{equation}
we just need to show how $|I_3|$ is obtained from $I_1$, $I_2$,
and $I_V=\det\mathbf{V}$, the three pieces of information locally
available to Alice. To achieve this goal we first note that a
direct calculation gives $I_V = I_1I_2 - I_4 + I_3^2$. Using
Eq.~(\ref{I4}) we see that $|I_3|$ follows from
$|I_3|^2 - 2 |I_3| \sqrt{I_1I_2} + I_1I_2 - I_V = 0.$
One of its
roots is not acceptable since it implies $\mathbf{V}< 0$.
Therefore, we are left with
\begin{equation}
|I_3|=\sqrt{I_1I_2} - \sqrt{I_V}. \label{I3}
\end{equation}
Hence, substituting Eqs.~(\ref{I4}) and (\ref{I3}) in
Eq.~(\ref{separabilidade}), Alice is able to unequivocally tell
whether or not she shares an entangled two-mode Gaussian state
with Bob.
Finally, if her state is entangled then $I_3<0$
\cite{simon} and, for a symmetric state ($I_1=I_2$), Alice can
quantify its entanglement via the entanglement of formation
($E_f$) \cite{Gie03,Rig04}:
\begin{equation}
E_f(\rho_{12}) = f\left(2 \sqrt{I_1+|I_3|-\sqrt{I_4 + 2 I_1
|I_3|}}\right), \label{ef}
\end{equation}
where $f(x)=c_+(x)\log_2(c_+(x)) - c_-(x)\log_2(c_-(x))$ and
$c_{\pm}(x)=(x^{-1/2}\pm x^{1/2})^2/4$. For arbitrary two-mode
Gaussian states ($I_1\neq I_2$) Alice can work with lower bounds
for $E_f$ \cite{Rig04} or calculate its negativity or logarithmic
negativity \cite{vidal}. This last two quantities are the best
entanglement quantifiers for non-symmetric two-mode Gaussian
states and are given as analytical functions \cite{adesso,adesso2}
of the four invariants here obtained from local measurements:
$I_1$, $I_2$, $|I_3|=\sqrt{I_1I_2} - \sqrt{I_V}$, and
$I_4=2|I_3|\sqrt{I_1I_2}$, with $I_V=\det\mathbf{V}$ given by
Eq.~(\ref{detV}). It is worth mentioning that $I_1$ ($I_2$)
can easily be determined by the measurement of the purity (Wigner
function at the origin of the phase space) of Alice's (Bob's) mode
alone \cite{fiurasek2,ban}. This measurement is less demanding
than the ones required to reconstruct ${\bf V}_1$ and 
$\mathbf{V}_2$ \cite{rigolin}.

Besides furnishing all the entanglement properties of an arbitrary
two-mode Gaussian state, the previous local protocol can also be
employed to reconstruct the covariance matrix for some particular
types of Gaussian states. To see this, let ${\bf\Gamma}_1$ be
explicitly written as
\be {\bf
\Gamma}_1=\left(
\begin{array}{cc}
\eta_1 + \frac{1}{2}& \mu_1 \\
\mu_1^* & \eta_1 + \frac{1}{2}\end{array} \right),
\ee
where
\br &\eta_1&=\langle a_1^\dagger a_1\rangle_{e}-\langle
a_1^\dagger a_1\rangle_{o}, \label{eta1}
 \\
\mu_1=\langle a_1^2\rangle_{e}&-&\langle a_1^2\rangle_{o}, \
\mu_1^*=\langle (a_1^\dagger)^2\rangle_{e}-\langle
(a_1^\dagger)^2\rangle_{o},  \label{mi1} \er
being $\langle \cdot \rangle_e$ and $\langle \cdot \rangle_o$ the
mean values for Alice's even and odd subensembles, respectively.
From this identity it is clear that ${\bf\Gamma}_1$ does not
necessarily represent a physical state since $\eta_1$ can take 
negative values \cite{haruna1}.
From Eq. (\ref{schur_rel1}) we obtain the following two relations,
\br n_1-\eta_1&=&\frac{1}{\left(n_2+\frac{1}{2}\right)^2-\vert
m_2\vert^2} \left\{\!\!\left(\vert m_c\vert^2\!\!+\vert
m_s\vert^2\right)\!\!\left(\!n_2\!+\!\frac{1}{2}\!\right)\right.\nonumber\\
&&\left. -2\Re e(m_2m_sm_c^*)\right\}, \label{eq1}\\
m_1-\mu_1&=&\frac{1}{\left(n_2+\frac{1}{2}\right)^2-\vert m_2\vert^2}
\left\{2m_sm_c\left(n_2+\frac{1}{2}\right)\right.\nonumber\\
&&\left.-m_2^*m_c^2-m_2m_s^2\right\}. \label{eq2} \er
Eqs.~(\ref{eq1}) and (\ref{eq2}) give the matrix elements of
$\mathbf{\Gamma_{1}}$ as a function of the matrix elements of
$\mathbf{V}$. If $m_c$ and $m_s$ are real (if either $m_c$ or
$m_s$ is zero) Eqs.~(\ref{eq1}) and (\ref{eq2}) can be inverted to
give $m_c$ and $m_s$ (either $m_s$ or $m_c$).

Let us explicitly solve the previous equations for an important
case, namely the ones in which $\textbf{C}\textbf{C}^\dagger=\vert
m_i\vert^2\textbf{I}$, where $i=c$ or $s$ and \textbf{I} is the
identity matrix. 
The states
comprehending this class are the ones where \textbf{C} has only
diagonal or non diagonal elements, i.e., $m_s=0$ and $m_c\neq0$ or
$m_c=0$ and $m_s\neq0$, reducing
the unknown quantities to two, namely the absolute value and the
phase of $m_s$ or $m_c$.
Remark that 
if $i=s$ the system is separable, since $\det\textbf{C}=\vert
m_s\vert^2\geq0$, i.e., the correlation between the two modes is
strictly classical \cite{simon}. Otherwise, if $i=c$ the state is
not necessarily separable, possibly being entangled, for in this
case $\det\textbf{C}=-\vert m_c\vert^2\leq0$. This last case is
more interesting since  it represents a class of
states that might show non-local  features \cite{simon}.

From  Eqs. (\ref{eq1}) and (\ref{eq2}) 
the diagonal (off-diagonal) elements of \textbf{C},
$m_i=\vert m_i\vert e^{i\phi_i}$, for $i=s$ ($i=c$), are
\br \vert
m_i\vert^2&=&\frac{(n_1-\eta_1)}{n_2+1/2}
\left[\left(n_2+ 1/2 \right)^2
-\vert m_2\vert^2\right], \label{mc}\\
e^{2i\phi_i}&=&\left(\frac{\mu_1-m_1}{n_1-\eta_1}\right)
\frac{n_2+ 1/2}{m_{2i}},
\er
where $m_{2c}=m_2^*$ and $m_{2s}=m_2$.
Note that whenever $m_2=0$, $\phi_i$ 
becomes undetermined. This problem can be solved by locally (unitary) 
transforming the two-mode squeezed state to a matrix
$V'_2$ with $m'_2\neq 0$, where $\phi'_i$ can be determined. Then,
transforming back, we get $\phi_i$. Fortunately,  there
are various experimentally available bipartite Gaussian states 
in which all the parameters are real, $m_s(m_c)=m_1=m_2=0$, and
$m_c(m_s)\neq 0$. For these states, Eq. (\ref{mc}) is sufficient 
to determine $\mathbf{C}$.

A natural and important example belonging to this class is the
two-mode thermal squeezed state 
\cite{daffer}, which is generated in a nonlinear crystal with
internal noise. Its covariance matrix is \be \textbf{V}=\left(
\begin{array}{cccc}
n+ \frac{1}{2} & 0 & 0 & m_c \\
0 & n+ \frac{1}{2} & m_c & 0 \\
0 & m_c & n+ \frac{1}{2} & 0 \\
m_c & 0 & 0 & n+ \frac{1}{2} \\ \end{array} \right),
\label{cov_esp} \ee
where $n$ and $m_c$ are time dependent functions having as
parameters the relaxation constant of the bath as well as the
nonlinearity of the crystal \cite{daffer}.
In this case the protocol involves only simple local measurements,
i.e., those to get $n$, $\langle a_1^\dagger a_1\rangle_{e}$ and
$\langle a_1^\dagger a_1\rangle_{o}$ (or equivalently $\eta_1$) by
Alice, and the parity measurements by Bob. The classical
communication corresponds to Bob informing Alice the instances he
performs the parity measurement in his mode and the respective
outcomes. Hence, Eq.~(\ref{mc}) reduces to
\be m_c^2=(n-\eta_1)\left(n+ 1/2\right). \label{mc_esp} \ee
Experimentally, $n$ and $\eta_1$ (Eq.~(\ref{eta1})) are readily
obtained by photodetection, while the parity measurement is
related to the determination of Bob's mode Wigner function at the
origin of the phase-space \cite{davidovich}, or alternatively to
his mode's purity, both of which can be measured by photocounting
experiments \cite{fiurasek2,ban}.

\begin{figure}[!ht]
\includegraphics[width=7cm]{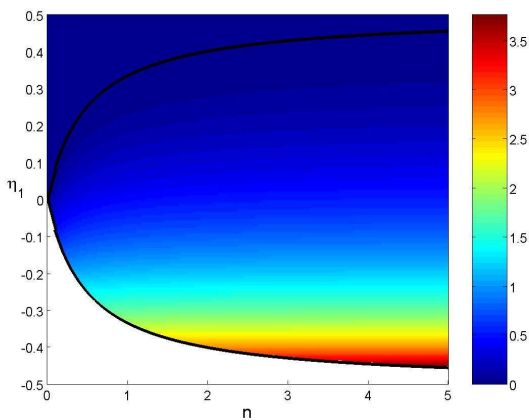}
\caption{\label{fig1}(Color online) Above the upper solid curve
lie the separable states. Below it, entanglement is quantified via
 $E_f$ (Eq.~(\ref{ef})) up to the lower
curve, where the pure entangled states are located. Below this
curve there
exist no physical states.}
\end{figure}

We can also study the P-representability \cite{footnote2} for 
the state (\ref{cov_esp}), which in this case is equivalent to the Simon
separability test \cite{simon,marcos1}. A
two-mode Gaussian state is P-representable iff
$\textbf{V}-\frac{1}{2}\textbf{I}\geq0$,
where \textbf{I} is the unity matrix of dimension $4$. Explicitly,
this separability condition 
 in terms of the elements
of (\ref{cov_esp}) is equivalent to $n\geq |m_c|$. From this
inequality and Eq.~(\ref{mc_esp}) we see that for a given $n$
there exists a bound for $\eta_1$ below which the states are
entangled (upper solid curve in Fig. \ref{fig1}):
\be -\frac{n/2}{n+1/2}\le\eta_1\le\frac{n/2}{n+1/2}. \label{ineq}
\ee
The left bound in Eq. (\ref{ineq}) (lower solid curve in Fig. 1) is 
a consequence of the uncertainty
principle, delimiting the set of all physical symmetric Gaussian states (SGS). 
This bound is marked by all the pure states and the upper curve bounds 
(from below) the subset of all separable (P-representable) 
states \cite{marcos1}.
Thus, for the SGS class, photon number measurements, before
and after Bob's parity measurements, are all Alice needs to
discover whether or not her mode is entangled with Bob's. 
The exquisite symmetry of those two antagonistic bounds is quite 
surprising, and possibly valid only for the SGS class. 
There is another interesting feature for the SGS set that should be emphasized.
Note that $\eta_1=0$ contains all the states where Bob has equal chances 
of getting even or odd outcomes for his parity measurements, 
delimiting two subsets (even and odd). The even subset contains all the 
states where Bob has greater probabilities of getting even outcomes 
while the odd subset contains all the states where he has greater 
probabilities of getting odd outcomes.
The entanglement for states belonging to the SGS can be quantified 
through $E_f$ (Eq. (\ref{ef})) as depicted by the color scale in 
Fig.~\ref{fig1}. It is remarkable that the most entangled states 
(including the pure ones) are concentrated in the $\eta_1<0$ odd subset.

In conclusion, we have presented the minimal set of local
operations and classical communication that allows one to quantify
the entanglement of an arbitrary two-mode Gaussian state. One
important step towards the derivation of this protocol was the
mathematical identity relating the two-mode covariance matrix
determinant to the product of two local quantities, namely the
determinants of the one-mode correlation matrix and its Schur
complement.
In addition, we have also shown that the Schur complement of one
of the modes' covariance matrix is obtained via a set of parity
measurements on the other one. We have also explicitly discussed
how the protocol works for a particular class of Gaussian 
states belonging to the SGS set.
Within this class, for states written in its standard form, we
have shown that only photon number measurements (made before and
after a parity measurement on the other mode) are needed to
completely characterize the state's entanglement.  

\begin{acknowledgments}
This work is supported by FAPESP and CNPq.
\end{acknowledgments}

\newpage
\appendix

\section{Erratum}

Eq. (4), on page 2, of [Phys. Rev. Lett. 98, 150501 (2007)] is not
so general as we have previously thought. However, our scheme does
not rely on it, as we show in what follows.

The experimental proposal we presented in our Letter allows one to
locally obtain the matrices $\mathbf{V_1}, \mathbf{V_2}$, and
$\mathbf{\Gamma_1}$, without assuming any particular form for the
covariance matrix $\mathbf{V}$. We now show that with these three
matrices we can determine the four invariants that completely
characterize the entanglement content of a two-mode Gaussian
state. The first two invariants are
\begin{equation}
I_1=\text{det}(\mathbf{V_1}), \hspace{1cm}
I_2=\text{det}(\mathbf{V_2}).
\end{equation}
The third one is calculated remembering that
%
$
\mathbf{\Gamma_1} = \mathbf{V_1} -
\mathbf{C}\mathbf{V_2^{-1}}\mathbf{C^{\dagger}}.
$
%
A simple algebra on the previous expression gives,
%
$ \mbox{det}\left( \mathbf{V_1} -
\mathbf{\Gamma_1}\right)=\mbox{det}(\mathbf{C})\mbox{det}(\mathbf{V_2^{-1}})
\mbox{det}(\mathbf{C^{\dagger}}). $
%
But
$\mbox{det}(\mathbf{C}) = \mbox{det}(\mathbf{C^{\dagger}}) = I_3$
and $\mbox{det}(\mathbf{V_2^{-1}}) = 1/\mbox{det}(\mathbf{V_2}) =
1/I_2$.
Hence
\begin{eqnarray}
|I_3|&=&\sqrt{I_2\,\mbox{det}\left( \mathbf{V_1} -
\mathbf{\Gamma_1}\right)}.
\end{eqnarray}
Furthermore, $\mathbf{\Gamma_1}$ satisfies another mathematical
indentity,
%
$
I_V=\text{det}\mathbf{V} =
\text{det}{\mathbf{V_2}}\text{det}{\mathbf{\Gamma_1}}.
$
%
Therefore, since we have $\mathbf{\Gamma_1}$ and $\mathbf{V_2}$,
we can also obtain $I_V$. But $I_V$ is related to the other four
invariants by the following expression,
%
$
I_V = I_1I_2 - I_4 + I_3^2.
$
%
Thus, the fourth invariant is simply
\begin{equation}
I_4 = I_1I_2 + I_3^2 - I_2\,\text{det}{\mathbf{\Gamma_1}}.
\end{equation}
Using $I_1$, $I_2$, $|I_3|$, and $I_4$, as obtained above with the
knowledge of $\mathbf{V_1}$, $\mathbf{V_2}$, and
$\mathbf{\Gamma_1}$, we can apply the Simon separability test (Eq.
(3) of our Letter). If a two-mode Gaussian state is entangled we
know for sure that $I_3<0$ and we can, therefore, fully quantify
its entanglement either via the entanglement of formation or the
negativity/logarithmic negativity, as discussed in our Letter.




Finally, we must emphasize that the main result of our Letter
remains unchanged: it is possible to completely characterize via
local operations and classical communication (LOCC) the
entanglement content of an arbitrary two-mode Gaussian state.
Furthermore, all the results presented in the Letter remain valid.

We want to thank Yang Yang, Fu-Li Li and Hong-Rong Li for also
calling our attention on the problems related to Eq. (4) of our
Letter while this erratum was being formulated.


\begin{thebibliography}{99}
\bibitem{comment1} We neglect first order moments since they can 
be set to zero locally without changing the entanglement content.
\bibitem{footnote2} A P-representable state has a non-negative 
probability distribution, being thus classical.
\end{thebibliography}
\end{document}